\documentclass[manuscript,screen]{acmart}

\usepackage{graphicx}

\AtBeginDocument{%
  }

\setcopyright{none}
\copyrightyear{2018}
\acmYear{2018}
\acmDOI{XXXXXXX.XXXXXXX}

\renewcommand\footnotetextcopyrightpermission[1]{}   
\begin{document}

\title{Generative AI \& Two Forms of Decoupling}


\author{Shira Gur-Arieh}
\email{sgurarieh@sjd.law.harvard.edu}
\affiliation{%
  \institution{Harvard University}
  \city{Cambridge}
  \state{Massachusetts}
  \country{USA}
}

\author{Sina Fazelpour}
\email{s.fazel-pour@northeastern.edu}
\affiliation{%
  \institution{Northeastern University}
  \city{Boston}
  \state{Massachusetts}
  \country{USA}
}

\renewcommand{\shortauthors}{Gur-Arieh et al.}

\begin{abstract}

Textual artifacts are sometimes valued not only for the words on the page, but for the human activity involved in producing them. The effort invested in a carefully tailored email can signal genuine interest; composing an apology might involve attending to another person’s hurt and deciding how to respond; and requiring a judge to give written reasons may induce more careful deliberation. AI models create what we call a \textit{decoupling problem}: they make it possible to produce text without undergoing the relevant human activity, severing its connection to values traditionally sustained by that activity. This Article develops a framework for understanding what decoupling puts at stake, and how institutions that use text to make consequential decisions may reshape their practices in response. \textit{First}, it offers a taxonomy of the grounds for valuing the production process, distinguishing whether that process matters for what it evidences, induces, or helps constitute. \textit{Second}, it argues that text can function as a kind of \textit{boundary object}, allowing institutions to rely on the same artifact without resolving disagreements about why the practice is valuable. AI can separate functions previously served by the same practice – a \textit{second-order decoupling} that brings unresolved questions about the practice’s purposes into view. \textit{Third}, it argues that institutional efforts to repair decoupling may recover some functions without preserving others. Such responses are likely to favor functions that are more legible, whose loss demands immediate attention, or whose stakeholders have greater influence. It therefore calls for more explicit deliberation about which particular functions of a practice to preserve, with meaningful representation for those whose interests might otherwise be overlooked.
  
\end{abstract}


\keywords{Generative AI, decoupling, process–artifact relationships, costly signaling, boundary object, textual artifacts, institutional design}

\maketitle

\section{Introduction}

In 2022, fans who had paid \$599 for “hand-signed” limited edition copies of Bob Dylan’s \textit{The Philosophy of Modern Song} began comparing photographs of their signatures online and noticed they were identical. Dylan, it emerged, had signed the books with an autopen – a machine that mechanically reproduces a signature from a template. The publisher offered refunds, and Dylan issued a rare public apology, calling his use of the machine “an error in judgment”. As Pegah Moradi and Karen Levy observe in their study of the autopen, the episode is puzzling only if one thinks an autograph is simply ink on a page \cite{moradi2024autopen}. After all, the machine-made signature was visually indistinguishable from the real one – if anything, more faithful to Dylan’s handwriting than a rushed scribble at a signing table. But, as the authors explain, signatures derive significance from the acts through which they are produced. Depending on the context, signing can express care, acknowledge responsibility, or give an object a particular authenticity and personal connection. The autopen can reproduce the signature’s physical form while bypassing the activity on which these values depend.

We call this dynamic \textit{decoupling}, and argue that Generative Artificial Intelligence (AI) extends it to a much wider range of textual artifacts.\footnote{By textual artifacts, we mean both those in formal and natural language. Because we focus on text in this Article, we concentrate mostly on Large Language Models (LLMs). We expect the same dynamic wherever a generative model can produce an artifact whose value depends on how it was made, including images, music, video, and physical objects. The autopen and the camera, discussed below, are pre-digital instances of the same structure.} Like a signature, text can derive value not only from the words on the page, but also from the human activity involved in producing it. A carefully tailored email from a prospective student, for example, tells you about their research and its connection to your own; but the time and effort involved in the process of preparing it also give you reason to take their interest in working with you seriously. Similarly, an apology or condolence note may offer comfort through its words, though part of its significance lies in the work of attending to another person’s hurt and deciding what to say. That expectation of personal attention may help explain why students were upset when Vanderbilt’s Peabody College sent a condolence email after the Michigan State University shooting, with a disclaimer noting it had been generated with the help of ChatGPT \cite{levine2023vanderbilt}. AI now makes it possible to produce a textual artifact without undergoing the process, decoupling it from some latent value(s) that traditionally made it meaningful in some sense.

A growing literature examines versions of this decoupling problem across otherwise disparate domains – from social-media posts and interpersonal apologies \cite{magnus2025chatbot, battisti2025secondperson, rubin2025empathy}, to hiring materials \cite{cui2025signaling, galdin2025making}, academic papers \cite{novelli2026second, hu2026undoing, treude2026artifact}, and legal documents \cite{re2024artificial, grimmelmann2026generative, meers2026aislop, dooling2026ghostwriting} – analyzing the phenomenon under various names and frameworks \cite{wojtowicz2025mentalproof, gallacher2026mirror}. Building on this work, the Article makes three contributions.

\textit{First}, it develops a taxonomy of process–artifact relationships to distinguish the different grounds for valuing the production process, clarifying what, exactly, decoupling puts at stake. A written judgment, for example, can provide evidence that a judge considered the parties’ arguments; composing it can induce deliberation; and its authoritative status depends on the legally authorized process through which it is issued. The Article also shows how the losses associated with these functions differ in who bears them and when they become apparent. \textit{Second}, the Article argues that text has historically functioned as a kind of boundary object \cite{StarGriesemer1989BoundaryObjects}, allowing institutions to rely on the same artifact without necessarily agreeing about which of these functions justified the practice. By separating artifact from process, AI can also unbundle those functions that previously traveled together – what we call \textit{second-order decoupling} – making these unresolved questions harder to avoid. \textit{Third}, it argues that institutional efforts to repair decoupling may restore some functions without preserving others. These responses are likely to favor functions whose loss is more immediately felt, whose value is more legible, or whose stakeholders have greater power to shape redesign. It therefore calls for explicit deliberation about which functions to preserve, with meaningful representation for those whose interests might otherwise be overlooked.

\section{Process-Artifact Relationships}
While this general account of decoupling laid out in the introduction – producing text while bypassing human activity on which some of its value traditionally depended – captures the core dynamic, it obscures the different functions served by the process in relation to the artifact, and, consequently, the different kinds of loss that arise when that connection is severed. Drawing on a range of theoretical traditions, this Article identifies three distinct roles that the underlying process of writing serves: it acts as evidence of hard-to-observe traits; it forms or induces skill, deliberation or knowledge; and it constitutes, or contributes towards, the status or significance of the artifact itself.

Consider, first, an \textit{evidentiary function}. Here, we value text because it reveals some otherwise hidden quality of its producer. A student essay may provide evidence of what the student knows; a job application may reveal the applicant’s interest, effort, or competence; a carefully tailored email from a student to a prospective advisor signals genuine familiarity with the professor’s work and desire to work with them. Scholars often describe this in costly signaling terms:  an artifact is informative because it is more costly to produce for those who lack the underlying trait \cite{spence1973job}. AI weakens this signal by reducing that cost differential, creating an inference problem: the text becomes less reliable evidence of what lies behind it.

Contrast this with a \textit{formative function}, in which the demand for a textual artifact \textit{induces} a valuable process. A student essay, for example, may be used as evidence of what a student knows, but it is often assigned because the act of writing actually develops that knowledge or skill. In other words, the difficulty of formulating and revising an argument is central to the assignment’s educational function. Similar examples arise outside the context of education: journaling, for example, fosters self-understanding, and requiring judges or peer reviewers to explain their reasoning (as opposed to simply stating the outcome) forces deliberation. This category is consistent with certain pedagogical approaches and with theories of “writing-as-thinking” \cite{emig1977writing}. Counter to the view that writing is merely the inert output of an already-completed internal thought, this framework recognizes that creating, reading, rearranging, and revising sentences are parts of the intellectual or cognitive process itself.

Lastly, consider a \textit{constitutive function}. While evidentiary and formative claims describe something that happens in the world (i.e., an inference becomes unreliable, or a capacity fails to develop), a constitutive claim concerns something somewhat subtler: how the process of producing a textual artifact forms part of what the artifact is, or gives it a particular status or significance. While AI promises to spare us the work of production, sometimes the effort, attention, or care invested in that work is precisely what gives an artifact a particular kind of meaning \cite{silbey2025aislop}. Part of what a fan values in Dylan’s autograph is possessing something he personally signed; the act of signing \textit{creates} a personal connection that reproduction alone cannot supply. The relevant object of value is an aggregate process-artifact unit. Consequently, two texts containing exactly the same words may differ in kind because they have different production histories. Several theoretical traditions capture versions of this idea. Gwen Bradford, for example, models such a process-artifact unit through her account of achievement: an achievement consists of a difficult process that competently culminates in a product, and comprises both the process and the product \cite{bradford2015achievement}. In one of her examples, two authors produce books of equal quality, yet one constitutes the greater achievement because it was more difficult to produce. Austin’s theory of speech acts provides another example. Whether certain words constitute a promise or an apology depends not only on the words but also on who utters them and under what circumstances \cite{austin1962how}. Constitutive relationships can also be fixed by social convention or institutional rule. Law, for example, specifies procedures through which text acquires legal effect: the exact same language constitutes a valid statute, will, or guilty plea only if it is produced according to the requisite institutional procedures \cite{fuller1941consideration, gulliver1941classification}.

The point is that when AI produces the text in place of a human, it detaches the artifact from something else we care about. The text becomes less reliable as evidence of an unobservable quality, the human process that writing would otherwise induce is bypassed, or the artifact loses a status or significance that depends on how it was produced. These functions can be contrasted with an \textit{output-oriented account}, on which a practice is valued strictly for the quality or usefulness of the artifact it yields. In other words, the production process matters only insofar as it contributes to that result. If an AI model can produce an equally satisfactory textual artifact, bypassing the human activity traditionally required to produce it, this does not, on this account, constitute a loss.

These categories describe broad types of function, each encompassing further distinctions. To say that a text serves an evidentiary function, for example, leaves open what it provides evidence \textit{of} – whether the producer’s knowledge, effort, interest, care, or something else. Similar fine-grained distinctions arise within the formative and constitutive categories. 

We also emphasize that the evidentiary, formative, and constitutive losses we describe are measured against what we call the "traditional process", by which we mean the way an artifact was produced before generative AI. This process may already have involved some kinds of tools ranging from search engines to programming environments. We recognize that, as AI changes the ways that people work, institutions may reconsider what they might want a text to evidence, or its production to cultivate or constitute;  they may shift, for example, toward assessing or developing competence in working \textit{with} AI. These expectations are thus open to revision, and a claim of a particular loss must explain why the displaced activity remains essential even once an automated alternative exists. In the constitutive case, this echoes debates over achievement and enhancement, which ask why some forms of assistance undermine what is considered an accomplishment while others are accepted (Suits illustrates the point with mountain climbing: ascending with ropes still counts as climbing, while arriving by helicopter does not \cite{suits1978grasshopper}). Revisions of this kind, however, call for explicit discussion of what a practice should achieve and why, which requires, at a minimum, making the new process legible enough to see what it involves and which functions it serves.

Crucially, the losses associated with these three functions differ in what triggers them, who bears their consequences, and when they become apparent. Evidentiary loss is often felt immediately by third-party evaluators, who can no longer draw the same inferences from a text. This can happen even when AI was not actually used: its availability alone can weaken the artifact’s value as evidence. Formative losses, by contrast, fall most directly on producers, particularly novices, and on institutions that are in charge of developing their capacities. Although formation may initially seem like a private benefit to the producer, the capacities it develops can also serve a wider community. Developing one's own scholarly judgment, for example, could potentially support peer review, mentorship, and the ability to recognize important research questions and detect mistakes. Still, these losses are not as visible because their consequences are dispersed and unfold over time. Constitutive losses have a more varied profile, but two kinds of interest are especially salient. Recipients may lose the relational significance an artifact derives from its production history, as with Dylan’s autograph or an apology letter. Because that underlying process does not typically leave a visible trace in the artifact, the loss may be recognized only later, when the history of its production becomes known. For producers, the interest lies in occupying and exercising a recognized role within a practice. Even a substitute that perfectly captures what someone would have contributed – for example, an LLM simulating the content of a citizen’s public comment – can deprive them of agency or voice in producing the artifact, or of a policy formed through their own participation. It may satisfy another actor’s need for an output while denying the person whose contribution it replaces an opportunity to take part.

\begin{figure}[ht]
    \centering
    \includegraphics[width=\textwidth]{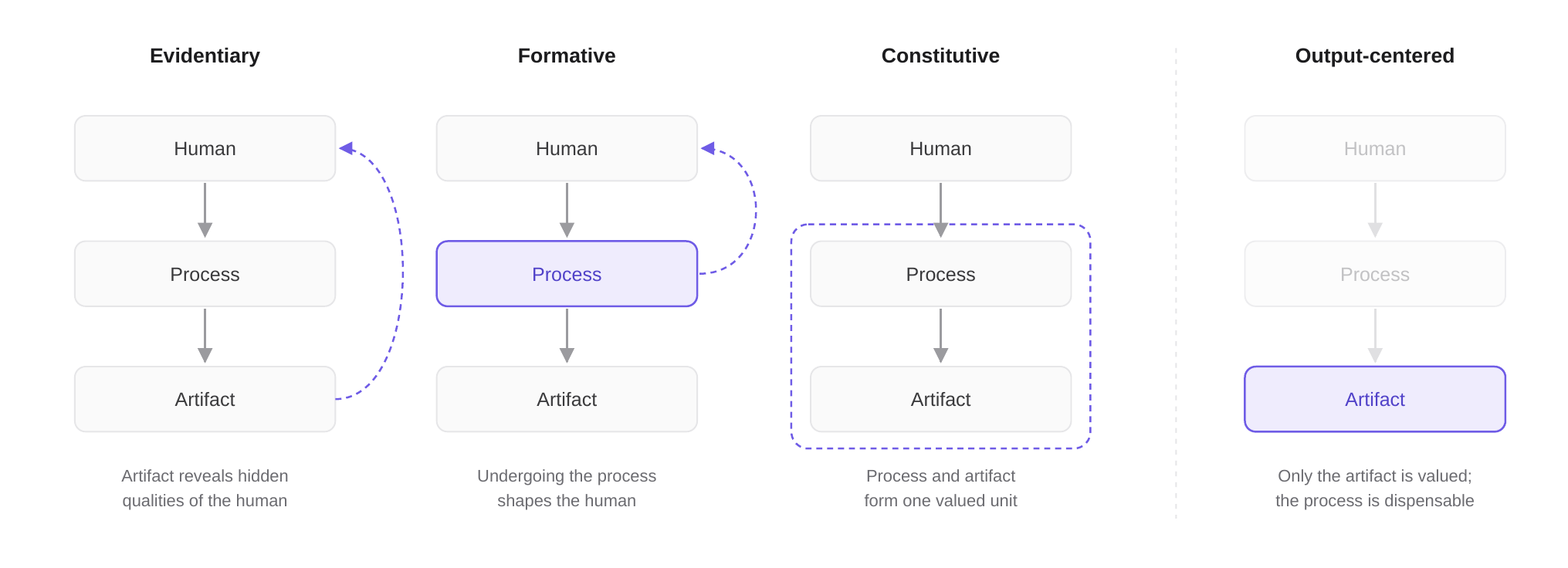}
    \caption{Taxonomy of process-artifact relationships.}
    \label{fig:process-artifact-taxonomy}
\end{figure}

\section{Text as a Boundary Object: Second-Order Decoupling}

Importantly, these categories are neither mutually exclusive nor inherent to the text itself. A single artifact can serve multiple functions simultaneously – a student essay, for instance, can be assigned to induce the cognitive process of learning while also serving as evidence of acquired knowledge. An apology letter might be valued as evidence of the sender’s care; because composing it induces genuine reflection in the sender; or because its production history is what gives it its status as an apology, as opposed to merely words in its shape. Moreover, the function we assign to a text is itself a normative choice, and institutional actors may differ in how they understand the artifact's value and its relationship to the process that produced it. Such disagreements can arise even \textit{within} a category: actors may agree that a text should serve an evidentiary function, for example, while disagreeing about which qualities of the producer it should reveal.

The student essay and apology letter are simple examples, but consider a slightly more complicated one with higher institutional stakes: the doctrine of “hard look” review, which requires administrative agencies to provide detailed, reasoned explanations for their regulations for courts to evaluate. This requirement can be justified under different kinds of theories \cite{stephenson2006costly}. Under one view, the mandate induces a valuable process: the arduous task of drafting forces the agency into genuine, careful deliberation. Under a different account, however, the requirement instead helps courts learn whether the agency actually values the policy it is defending. A thorough record is costly to produce, so an agency’s willingness to incur that cost can serve as evidence that it places a high value on the policy. This, in turn, allows courts to treat visible effort as evidence of the agency’s underlying commitment. Notice that these two accounts assign very different roles to the production process itself: on the first, the effortful process is a desirable feature – indeed, the point – indispensable because undertaking it generates the deliberation we care about; on the second, the process is an expenditure, valuable only insofar as it reveals information about the agency’s underlying commitment.

In the ordinary course of institutional life, these competing views typically remain implicit. We might say that a text functions as a kind of “boundary object”: a shared artifact that allows different actors to rely on the same text for distinct purposes without necessarily needing to agree on exactly why it is valuable \cite{StarGriesemer1989BoundaryObjects}.\footnote{For a similar, yet slightly different kind of conceptualization of this idea in the context of mathematics, see Terence Tao, Mathematics in the Age of AI \cite{tao2026mathematics}. Tao argues that the goals of mathematics have historically been positively correlated – progress on one moved you toward the others – and that because of this correlation, “one could use one or two of these goals as convenient proxies for the others, and leave the remainder implicitly stated at most”. Under excessive optimization, the previously aligned goals “diverge from one another”, and what’s being stress-tested is “the largely implicit framework of mathematical values and practices […] what we consider a contribution to be, what we reward […] and who – or what – we regard as having done the work”.}  But an interesting, perhaps overlooked feature of technological shortcuts like AI is that, by creating a new way to reach the artifact without the process, such shortcuts can also unsettle alignment among functions previously supported through the same practice. We call this dynamic \textit{second-order decoupling}. This can force us to articulate a particular theory of what aspects of the underlying practice we actually care about.

To make this idea concrete, consider a few real-world, pre-LLM examples. In 2011, while abroad, President Obama authorized an autopen to sign legislation. The episode pitted two accounts of what a presidential signature does. On an evidentiary account, the signature simply authenticates a decision the President has already made, making the physical act dispensable. On a constitutive account, on the other hand, personally signing is itself part of the prescribed procedure through which a bill becomes law. The White House relied on a 2005 Office of Legal Counsel opinion which adopted the former view, concluding that the President could delegate the inscription, but not the underlying decision \cite{nielson2005autopen}. Thus, the autopen separated authorization from personal execution, a distinction handwriting had obscured.

Another illustrative example comes from the FCC’s 2017 net-neutrality rulemaking, when the agency’s docket was flooded with millions of automated comments. Like the autopen episode, this event raised deeper questions about the purpose of the practice of notice-and-comment \cite{herz2020malattributed}. If public comments matter primarily for the information and arguments they provide to the agency, then automation is, aside from the logistical burden it creates, not necessarily a problem: what matters is the content of the submission that  feeds into the rulemaking process. But if a comment also serves as evidence that a citizen cared enough to invest the time and effort to participate, automation changes its significance. Scripts made it possible to submit comments in citizens’ names without the underlying act of participation that the comment had previously evidenced. Here too, the technological shortcut pulled apart two understandings of the same practice, and forced into the open a disagreement that the shared artifact previously allowed institutions to leave unresolved. 

While this dynamic is not unique to AI – other historical technological shortcuts (autopen, deepfakes, boilerplate contracts, and the camera, to name a few) can be described through the same framework – our claim is that institutions that rely on text are about to experience it on a vast scale. The distinction between the two levels of decoupling developed throughout the preceding discussion can now be stated more formally:

\smallskip
\noindent
\begin{tabular}{@{}lp{0.70\linewidth}@{}}
\textbf{First-order decoupling} & occurs when a textual artifact can be produced without the human process traditionally required to create it, severing its connection to what that process evidences, induces, or constitutes. \\[4pt]
\textbf{Second-order decoupling} & occurs when first-order decoupling disrupts the alignment among a textual practice's functions that previously allowed actors to rely on the same artifact without agreeing on what justified the practice. \\

\end{tabular}
\smallskip

\vspace{-0.5em}

\section{The Normative Stakes of Institutional Repair}

When AI separates the artifact from its traditional production process (first-order decoupling), institutions will likely redesign the practice to restore what was lost. But where functions previously accommodated by the same practice are no longer aligned (second-order decoupling), a replacement designed to restore one function need not preserve the others. For example, when AI enables students to bypass the work of writing an essay, an oral exam may bring back evidence of understanding without preserving the practice of constructing and revising an argument; a Socratic chatbot, on the other hand, might support learning, but it will not help in establishing what students can do independently, without AI. Or consider a more complex example of a PhD thesis, which arguably serves all three functions. Alongside its evidentiary and formative roles, it can also constitute a kind of achievement whose significance depends partly on the candidate’s own intellectual contribution. Responding to AI's growing ability to generate proofs – and the resulting weakening of the link between mathematical results and their authors’ understanding – one recent proposal would award PhDs primarily on demonstrated expertise, even when AI produced the results \cite{litt2026beginning}. Students would still learn under supervision and demonstrate understanding through rigorous defenses, but would not need to have originated those results themselves. The proposal thus seeks to preserve evidentiary and formative functions while relinquishing – or at least redefining – the thesis’s constitutive significance as the candidate’s own intellectual achievement.

Selective preservation of this kind is not, on its face, a bad thing. Second-order decoupling can prompt institutions to confront questions that the old practice left unresolved and make explicit choices about what should be preserved. But we think the functions at stake do not have equal prospects of preservation. The asymmetries described in Section 2 give us reason to expect that redesign will favor functions whose value is more institutionally legible, whose loss is more immediately felt, or whose stakeholders have greater power and resources to secure a response. Which functions survive may therefore reflect these asymmetries more than a considered judgment about what the practice should achieve.

First, we expect institutional responses to decoupling to be shaped disproportionately by evaluators’ interests and incentives.\footnote{We expect output-oriented concerns to take priority over all three functions in many settings, especially where powerful actors care chiefly about what the artifact can accomplish, regardless of its production history. Among the three process-dependent functions, however, differences in when losses are felt and who can press for a response make evidentiary repair a likely institutional priority.} For third-party evaluators whose work involves reading and assessing textual artifacts – such as teachers, employers, editors, and judges – decoupling creates an immediate practical problem: they must continue making decisions even though a text no longer supports the inferences on which those decisions once rested. This gives the evidentiary framing a particular urgency. Perhaps unsurprisingly, then, much of the literature and broader institutional discourse has approached decoupling through a predominantly evidentiary lens, often grounded more specifically in costly signaling theory. In its costly-signaling form, this framing treats the production process as a burden which is justified only by what it reveals, rather than an activity with value of its own. Repairs guided by this view may therefore seek another source of evidence of the relevant qualities, as in the oral exam example.\footnote{Some institutions are already moving in these directions. Employers place less weight on cover letters or take-home written exercises and instead rely on interviews, live assessments, or work trials \cite{cui2025signaling, altchek2026coverletters}; educators are replacing written assignments with oral examinations \cite{gecker2026oralexams}; and at least one newsroom has responded to the influx of AI-generated written pitches by requiring reporters to pitch stories by phone \cite{tameez2026deluge}.} But a replacement that supplies reliable evidence will not necessarily induce the activity that the original practice required. Returning to the “hard look” review example, an evidentiary framing might lead courts to place greater weight on the political capital an agency has invested in promoting a regulation \cite{stephenson2006costly}. That investment could signal how much the agency values the policy without inducing the deliberative work involved in drafting a reasoned justification. Where evaluators have less direct stake in the practice’s other functions, restoring the evidence they need may be enough for institutions to consider the problem “solved”, even if formative or constitutive losses remain.

Second, we expect institutional pressures to favor \textit{redescription}: institutions may come to understand a plural practice in purely output-oriented terms, reducing it to the delivery of its most visible output. Institutions have practical reasons to favor simplified, legible accounts of complex activities \cite{scott1998seeing}. Such accounts can, in turn, come to define what participants regard as valuable, displacing richer understandings of what the practice is for \cite{nguyen2024value}. The dynamic also has roots in the longer history of automation. Writing in 1984, Albert Borgmann observed that modern technology characteristically separates a desired commodity from the broader practice through which it was once obtained \cite{borgmann1984technology}. His canonical example is the replacement of the hearth by central heating. The hearth was what Borgmann called a “focal thing”, sustained by a web of practices: chopping and carrying wood, tending the fire, gathering around its warmth. Once warmth becomes independently available, tending the hearth can appear to have been merely an inefficient means of producing it (“we are inclined to think of these additional elements as burdensome”). Extending this analogy to our context, AI might also invite a similar redescription dynamic, where institutions – especially those that stand to benefit from automation – may begin to treat the text as the entire purpose of the practice. To illustrate, consider the example of clinical note-writing. Clinical note-taking is easily characterized as merely “documentation”: an administrative task whose purpose is to produce an accurate record. AI scribes are now being deployed across hospitals on the promise that they can produce those records more efficiently. But as recent accounts argue, particularly for trainees, writing the note often scaffolds clinical reasoning, forcing physicians to synthesize fragmented information and commit to an assessment \cite{abernethy2026scribes}. Once the practice is defined entirely by the record it yields, this formative value of note-writing is lost.

Redescription may also grow out of the difficulty of repairing decoupling. Institutions may still require authors to do the relevant intellectual work, but they can no longer rely on the finished text to establish that they have. Checking may require examining drafts, asking authors to explain their reasoning, enforcing AI bans – additional demands that are costly and still leave uncertainty. Institutions may therefore retreat to evaluating the output alone. A peer-reviewed journal, for example, frustrated by the difficulty of determining how much of a paper was AI-generated, may eventually decide to evaluate the submission solely on its final novelty and  “contribution to the field”. What begins as an epistemic problem – we can no longer tell whether the relevant process occurred – hardens into a normative claim – the process need not occur, so long as the resulting artifact is good enough.

\section{The Path Forward}

The examples in this Article have been deliberately  simple. In complex domains, however, the relationship between process and artifact is entangled with deeper disagreements that cut to the core of a discipline’s purpose. Academic scholarship, legal drafting, democratic participation, scientific research, journalism: these are thick practices, describable through very different accounts of how the process relates to the artifact, and which account holds depends on substantive views about what the practice just is. Our taxonomy takes a first step in helping articulate these disagreements, but deciding which elements of a practice must survive decoupling ultimately requires sustained, normative engagement from the communities themselves.

Mathematics offers a particularly revealing example of a community addressing such questions explicitly. Math might initially seem like an obvious case for judging intellectual work by its results alone: a proof's validity does not depend on who wrote it or how it was produced. Yet a claimed AI-generated solution to a longstanding Millennium Prize Problem has recently intensified debates about the broader goals of mathematical research. A recent statement signed by more than two dozen Fields Medalists reacting to these events questions whether an increasing supply of solutions will sustain the training and intellectual exchange that the work of finding them traditionally supported \cite{mathai2026misalignment}. Solving problems, they wrote, is “only a tool and proxy for achieving the primary goal of conceptual understanding and insight”. The earlier Leiden Declaration on Artificial Intelligence and Mathematics similarly identifies the cultivation of human understanding, clarity, and judgment as a purpose of research alongside producing results, and emphasizes that such expertise is essential to formulating significant new questions \cite{leiden2026declaration}. Even though these statements leave much open about how mathematical institutions should adapt, they make the field’s different purposes a subject of collective deliberation. As Terence Tao observed, these were questions “we should have been systematically discussing years ago” \cite{leiden2026declaration}. 

Fields Medalists, of course, lend considerable authority to these concerns; elsewhere, those who bear formative or constitutive losses may have less influence. We hope to encourage comparable conversations across some of the practices considered here. Such discussions may lead institutions to abandon particular inherited requirements or find new ways to sustain what they value. But making a practice’s purposes explicit is only part of the task: those with a stake in its less visible functions must also have a meaningful role in deciding its future.

\bibliographystyle{ACM-Reference-Format}
\bibliography{references}

@article{moradi2024autopen,
  author  = {Pegah Moradi and Karen Levy},
  title   = {{``A Fountain Pen Come to Life'': The Anxieties of the Autopen}},
  journal = {International Journal of Communication},
  volume  = {18},
  year    = {2024},
  pages   = {784--792},
  url     = {https://ijoc.org/index.php/ijoc/article/view/21842}
}

@article{stephenson2006costly,
  author  = {Matthew C. Stephenson},
  title   = {A Costly Signaling Theory of {``Hard Look''} Judicial Review},
  journal = {Administrative Law Review},
  volume  = {58},
  number  = {4},
  year    = {2006},
  pages   = {753--814},
  url     = {https://www.jstor.org/stable/40711887}
}

@article{tao2026mathematics,
  author  = {Terence Tao},
  title   = {Mathematics in the Age of {AI}},
  year    = {2026},
  journal = {arXiv preprint arXiv:2608.16753},
  doi     = {10.48550/arXiv.2608.16753},
  url     = {https://arxiv.org/abs/2608.16753}
}

@book{borgmann1984technology,
  author    = {Albert Borgmann},
  title     = {Technology and the Character of Contemporary Life: A Philosophical Inquiry},
  publisher = {University of Chicago Press},
  address   = {Chicago},
  year      = {1984},
  isbn      = {9780226066295}
}

@misc{mathai2026misalignment,
  title        = {A Severe Misalignment of {AI} in Mathematics},
  author       = {{Math and AI}},
  year         = {2026},
  month        = sep,
  note         = {Declaration published September 11, 2026},
  doi          = {10.5281/zenodo.22737750},
  url          = {https://www.mathandai.org/}
}

@misc{leiden2026declaration,
  author = {Jarod Alper et al.},
  title  = {Leiden Declaration on Artificial Intelligence and Mathematics},
  year   = {2026},
  month  = jun,
  doi    = {10.5281/zenodo.20302944},
  url    = {https://leidendeclaration.ai/}
}

@inproceedings{wojtowicz2025mentalproof,
  author    = {Zachary Wojtowicz and Simon DeDeo},
  title     = {Undermining Mental Proof: How {AI} Can Make Cooperation Harder by Making Thinking Easier},
  booktitle = {Proceedings of the AAAI Conference on Artificial Intelligence},
  volume    = {39},
  number    = {2},
  pages     = {1592--1600},
  year      = {2025},
  doi       = {10.1609/aaai.v39i2.32151},
  url       = {https://ojs.aaai.org/index.php/AAAI/article/view/32151}
}

@book{bradford2015achievement,
  author    = {Gwen Bradford},
  title     = {Achievement},
  publisher = {Oxford University Press},
  year      = {2015},
  doi       = {10.1093/acprof:oso/9780198714026.001.0001},
  isbn      = {9780198714026}
}

@article{spence1973job,
  author  = {Michael Spence},
  title   = {Job Market Signaling},
  journal = {The Quarterly Journal of Economics},
  volume  = {87},
  number  = {3},
  year    = {1973},
  pages   = {355--374},
  doi     = {10.2307/1882010}
}

@article{emig1977writing,
  author  = {Janet Emig},
  title   = {Writing as a Mode of Learning},
  journal = {College Composition and Communication},
  volume  = {28},
  number  = {2},
  year    = {1977},
  pages   = {122--128},
  doi     = {10.2307/356095}
}

@techreport{nielson2005autopen,
  author      = {Howard C. Nielson Jr.},
  title       = {Whether the President May Sign a Bill by Directing That His Signature Be Affixed to It},
  institution = {Office of Legal Counsel, U.S. Department of Justice},
  year        = {2005},
  month       = jul,
  number      = {29 Op. O.L.C. 97},
  url         = {https://www.justice.gov/olc/opinion/whether-president-may-sign-bill-directing-his-signature-be-affixed-it}
}

@misc{litt2026beginning,
  author = {Daniel Litt},
  title  = {A Beginning for Mathematics},
  year   = {2026},
  month  = sep,
  day    = {14},
  url    = {https://proofsandprompts.com/2026/09/14/a-beginning-for-mathematics/},
  note   = {Proofs and Prompts}
}

@article{abernethy2026scribes,
  author  = {Jane Abernethy and Anna Shah and Belinda Chen and Stasia Reynolds and Scott M. Wright and Paul O'Rourke},
  title   = {Integrating {AI} Scribes into Medical Education: Guardrails for Preserving Clinical Reasoning},
  journal = {Journal of General Internal Medicine},
  volume  = {41},
  number  = {9},
  pages   = {2598--2602},
  year    = {2026},
  doi     = {10.1007/s11606-025-10149-w}
}

@book{suits1978grasshopper,
  author    = {Suits, Bernard},
  title     = {The Grasshopper: Games, Life and Utopia},
  year      = {1978},
  publisher = {University of Toronto Press},
  address   = {Toronto},
  isbn      = {9780802067449},
  doi       = {10.3138/j.ctvcj2w4h}
}

@article{levine2023vanderbilt,
  author  = {Levine, Sam},
  title   = {Vanderbilt Apologizes for Using ChatGPT in Email on Michigan Shooting},
  journal = {The Guardian},
  year    = {2023},
  month   = feb,
  day     = {22},
  url     = {https://www.theguardian.com/us-news/2023/feb/22/vanderbilt-chatgpt-ai-michigan-shooting-email}
}

@article{magnus2025chatbot,
  author  = {Magnus, P. D. and Buccella, Alessandra and D'Cruz, Jason},
  title   = {Chatbot Apologies: Beyond Bullshit},
  journal = {AI and Ethics},
  year    = {2025},
  volume  = {5},
  number  = {5},
  pages   = {5517--5525},
  doi     = {10.1007/s43681-025-00800-x}
}

@article{cui2025signaling,
  author  = {Cui, Jingyi and Dias, Gabriel and Ye, Justin},
  title   = {Signaling in the Age of AI: Evidence from Cover Letters},
  journal = {arXiv preprint arXiv:2509.25054},
  year    = {2025},
  doi     = {10.48550/arXiv.2509.25054},
  url     = {https://arxiv.org/abs/2509.25054}
}

@article{galdin2025making,
  author  = {Galdin, Anais and Silbert, Jesse},
  title   = {Making Talk Cheap: Generative AI and Labor Market Signaling},
  journal = {arXiv preprint arXiv:2511.08785},
  year    = {2025},
  doi     = {10.48550/arXiv.2511.08785},
  url     = {https://arxiv.org/abs/2511.08785}
}

@article{re2024artificial,
  author  = {Re, Richard M.},
  title   = {Artificial Authorship and Judicial Opinions},
  journal = {The George Washington Law Review},
  year    = {2024},
  volume  = {92},
  number  = {6},
  pages   = {1558--1590}
}

@article{grimmelmann2026generative,
  author  = {Grimmelmann, James and Sobel, Benjamin L. W. and Stein, David},
  title   = {Generative Misinterpretation},
  journal = {Harvard Journal on Legislation},
  year    = {2026},
  volume  = {63},
  number  = {1},
  pages   = {229--308},
  url     = {https://journals.law.harvard.edu/jol/2026/01/24/generative-misinterpretation/}
}

@article{meers2026aislop,
  author  = {Meers, Jed},
  title   = {{'Chucking Their Stuff into ChatGPT and Hitting Send': AI Slop in the Administrative State}},
  year    = {2026},
  journal = {SSRN Electronic Journal},
  doi     = {10.2139/ssrn.6898138},
  url     = {https://ssrn.com/abstract=6898138}
}

@article{novelli2026second,
  author  = {Novelli, Claudio and Floridi, Luciano},
  title   = {What is Left for Us? Second Scholarship Against the Degradation of Research by AI},
  journal = {arXiv preprint arXiv:2607.04049},
  year    = {2026},
  doi     = {10.48550/arXiv.2607.04049},
  url     = {https://arxiv.org/abs/2607.04049}
}

@article{hu2026undoing,
  author  = {Hu, Lily},
  title   = {Un-Doing Philosophy},
  journal = {Berlin Review},
  year    = {2026},
  number  = {21},
  month   = jul,
  day     = {10},
  url     = {https://blnreview.de/en/ausgaben/2026-08/als-ob-philosophieren-lily-hu}
}

@article{treude2026artifact,
  author  = {Treude, Christoph and Poskitt, Christopher M. and Hoda, Rashina},
  title   = {Rethinking Artifact Evaluation for Software Engineering in the Age of Generative AI},
  journal = {arXiv preprint arXiv:2604.16306},
  year    = {2026},
  doi     = {10.48550/arXiv.2604.16306},
  url     = {https://arxiv.org/abs/2604.16306}
}

@article{dooling2026ghostwriting,
  author  = {Dooling, Bridget C. E.},
  title   = {Ghostwriting the Government},
  journal = {Marquette Law Review},
  year    = {2026},
  volume  = {109},
  pages   = {767--821},
  url     = {https://scholarship.law.marquette.edu/mulr/vol109/iss2/7}
}

@article{battisti2025secondperson,
  author  = {Battisti, Davide},
  title   = {Second-Person Authenticity and the Mediating Role of AI: A Moral Challenge for Human-to-Human Relationships?},
  journal = {Philosophy \& Technology},
  year    = {2025},
  volume  = {38},
  pages   = {28},
  doi     = {10.1007/s13347-025-00857-w},
  url     = {https://link.springer.com/article/10.1007/s13347-025-00857-w}
}

@article{rubin2025empathy,
  author  = {Rubin, Matan and Li, Joanna Z. and Zimmerman, Federico and Ong, Desmond C. and Goldenberg, Amit and Perry, Anat},
  title   = {Comparing the Value of Perceived Human Versus AI-Generated Empathy},
  journal = {Nature Human Behaviour},
  year    = {2025},
  volume  = {9},
  number  = {11},
  pages   = {2345--2359},
  doi     = {10.1038/s41562-025-02247-w},
  url     = {https://doi.org/10.1038/s41562-025-02247-w}
}

@article{gallacher2026mirror,
  author  = {Gallacher, Paul},
  title   = {The Mirror Effect: How Generative AI Exposes the Proxy Architecture of Institutional Accountability},
  year    = {2026},
  journal = {SSRN Electronic Journal},
  doi     = {10.2139/ssrn.6415898},
  url     = {https://ssrn.com/abstract=6415898}
}

@article{StarGriesemer1989BoundaryObjects,
  author  = {Star, Susan Leigh and Griesemer, James R.},
  title   = {Institutional Ecology, `Translations' and Boundary Objects: Amateurs and Professionals in Berkeley's Museum of Vertebrate Zoology, 1907--39},
  journal = {Social Studies of Science},
  year    = {1989},
  volume  = {19},
  number  = {3},
  pages   = {387--420},
  doi     = {10.1177/030631289019003001}
}

@book{austin1962how,
  author    = {Austin, J. L.},
  title     = {How to Do Things with Words},
  editor    = {Urmson, J. O.},
  year      = {1962},
  publisher = {Clarendon Press},
  address   = {Oxford}
}

@article{fuller1941consideration,
  author  = {Fuller, Lon L.},
  title   = {Consideration and Form},
  journal = {Columbia Law Review},
  year    = {1941},
  volume  = {41},
  number  = {5},
  pages   = {799--824},
  doi     = {10.2307/1117840},
  url     = {https://www.jstor.org/stable/1117840}
}

@article{herz2020malattributed,
  author  = {Herz, Michael},
  title   = {Fraudulent Malattributed Comments in Agency Rulemaking},
  journal = {Cardozo Law Review},
  year    = {2020},
  volume  = {42},
  number  = {1},
  pages   = {1--67},
  url     = {https://cardozolawreview.com/wp-content/uploads/2021/01/42.1.1.Herz_.pdf}
}

@article{tameez2026deluge,
  author  = {Tameez, Hanaa'},
  title   = {Facing a ``Deluge'' from AI, This Publication Will Only Take Pitches by Phone},
  journal = {Nieman Journalism Lab},
  year    = {2026},
  month   = aug,
  day     = {18},
  url     = {https://www.niemanlab.org/2026/08/facing-a-deluge-from-ai-this-publication-will-only-take-pitches-by-phone/}
}

@article{altchek2026coverletters,
  author  = {Altchek, Ana},
  title   = {RIP Cover Letters: AI Is Killing One of the Most Annoying Parts of Getting Hired. But How Do You Stand Out Now?},
  journal = {Business Insider},
  year    = {2026},
  month   = jun,
  day     = {1},
  url     = {https://www.businessinsider.com/rip-cover-letters-generative-ai-hiring-2026-6}
}

@article{gecker2026oralexams,
  author  = {Gecker, Jocelyn},
  title   = {Perfect Homework, Blank Stares: Why Colleges Are Turning to Oral Exams to Combat AI},
  journal = {Associated Press},
  year    = {2026},
  month   = mar,
  day     = {25},
  url     = {https://www.ap.org/news-highlights/spotlights/2026/perfect-homework-blank-stares-why-colleges-are-turning-to-oral-exams-to-combat-ai/}
}

@article{gulliver1941classification,
  author  = {Gulliver, Ashbel G. and Tilson, Catherine J.},
  title   = {Classification of Gratuitous Transfers},
  journal = {Yale Law Journal},
  year    = {1941},
  volume  = {51},
  number  = {1},
  pages   = {1--39}
}

@book{scott1998seeing,
  author    = {James C. Scott},
  title     = {Seeing Like a State: How Certain Schemes to Improve the Human Condition Have Failed},
  year      = {1998},
  publisher = {Yale University Press},
  address   = {New Haven, CT}
}

@article{nguyen2024value,
  author  = {C. Thi Nguyen},
  title   = {Value Capture},
  journal = {Journal of Ethics and Social Philosophy},
  volume  = {27},
  number  = {3},
  pages   = {469--504},
  year    = {2024},
  doi     = {10.26556/jesp.v27i3.3048}
}

@article{silbey2025aislop,
  author  = {Jessica M. Silbey and Woodrow Hartzog},
  title   = {AI Slop},
  year    = {2025},
  journal = {SSRN Electronic Journal},
  url     = {https://ssrn.com/abstract=5870742},
  note    = {Last revised September 6, 2026}
}

\end{document}